\documentclass[aps,rmp,twoside,twocolumn]{revtex4}
\usepackage[latin1]{inputenc}
\bibpunct{[}{]}{;}{a}{,}{,}

\makeatletter

\providecommand{\LyX}{L\kern-.1667em\lower.25em\hbox{Y}\kern-.125emX\@}
\newcommand{\noun}[1]{\textsc{#1}}

\usepackage{pslatex,amssymb,citesort}

\makeatother

\begin{document}

\title{\Huge Are quantum `irreality' and `nonlocality' ineluctable?}

\author{A. F. Kracklauer\\Bauhaus Universit\"at; Weimar, Germany}

\begin{abstract}
The early history of the development of Quantum Mechanics is surveyed to discern
the arguments leading to the introduction of the notions of `irreal' wave functions
and `nonlocal' correlations. It is argued that the assumption that Quantum Mechanics
is `complete', i.e., not just a variant of Statistical Mechanics, is the feature
compelling the introduction of these otherwise problematic properties. Additionally,
a consequence of the error first found by \noun{Jaynes} in proofs of \noun{Bell}'s
``theorem'', is illustrated. Finally, speculation on the practical consequences
of recognising that ``entanglement'' is a feature of all hyperbolic differential
equations is proposed. 

\noindent{\bf Keywords}: irreality, nonlocality, entanglement, wave function
collapse
\end{abstract}
\maketitle

\section{History}

Twentieth century physicists faced extraordinary challenges in terms of the
scale of the phenomena to be explained. The extremely small sizes of the objects
covered by Quantum Mechanics (QM) and the very large scale of events covered
by Relativity, posed situations that were unimaginable within the then customary
understanding and concepts of science. This fostered, out of desperation, license
to introduce theretofore unacceptably exotic hypotheses, (e.g., quantisation,
frame independent light velocity) for which there was only indirect laboratory
evidence. At the same time, at least one desiderium was assumed, largely without
deep reflection, namely, that the new theories under development were fundamental
and complete (that is, that at their level they are theories of individual entities,
not theories quantifying statistics of ensembles of such entities).

Herein, I shall argue that the last mentioned assumption, \emph{completeness},
is the key underlying cause that the exotic (and arguably antirational) notions:
\emph{irreality} and \emph{nonlocality,} have been ensconced in QM. The reasoning
that led to this situation, was not concerned in the first instance with the
philosophically problematic nature of these features, indeed the terms themselves,
as well as their accepted technical denotation, appeared in the literature up
to years later than the introduction of the mathematical structure to which
they refer. 

For the sake of expository efficiency, herein a reconstructed line of reasoning
will be described that, I believe, is a composite of what happened, but not
in the mind of any one person. The actual development, to the extent that the
historically true story can be discerned at all in retrospect, was focused on
finding the mathematical structure that mimicked those aspects of the studied
phenomena accessible to experiment, which were nearly always imagined to be
just a portion of what happens `down deep'. In seeking these models or paradigms,
however, it is clear that the founding fathers individually quite early settled
on a personally preferred paradigm, and tilted their analysis to support it.
For QM, the accumulated effect, warped by sociological factors, is what has
become known as the ``Copenhagen interpretation''.

Some may be inclined to write off these concerns as `mere philosophy' of little
interest ``during working hours'' to physical scientists. However, one can observe
that, wherever there is a `philosophical' issue, the mathematics in use is also
afflicted with one or another pathology. Such a coincidence makes perfect sense
actually; mathematical problems (inconsistent or incomplete calculations) nearly
always parallel verbal problems (erroneous syntax) insofar as mathematics is
just the use of symbols and formalised algorithms as a sort of shorthand for
ideas originally expressed verbally. Indeed, all mathematics is taught by means
of oral explanations. Philosophical problems, following this logic, are often,
therefore, symptoms of physics problems. Moreover, there may well be very practical
consequences with regard to eventual applications derived from self consistent
interpretations.

\section{Irreality}

Irreal wave functions are of the form:
\begin{equation}
\label{1}
\psi (r_{1},r_{2})=\varphi (r_{1},r_{2})+\chi (r_{1},r_{2}),
\end{equation}
 where \( \psi (r_{1},r_{2}) \) represents a wave function for a combined system
of subsystems, and both \( \varphi (r_{1},r_{2}) \) and \( \chi (r_{1},r_{2}) \)
are the wave functions for potential outcomes. Wave functions of this form are
\emph{irreal} if the summands are logically mutually exclusive, i.e., states
that by all logic cannot exist simultaneously. There is, in addition, a continuous
variant of this same structure, in which, for example a point particle, which
can be at only one location at once, is represented by a wave packet finite
over several locations. All the conceptual features of irreality are evident,
nevertheless, in the binary version as captured in Eq. (\ref{1}); thus, let
us focus on it.

A prototypical example of a binary irreal wave function is the singlet state
used to describe the emission of correlated photon pairs for an EPR experiment:
\begin{equation}
\label{2}
\psi (1,2)=\frac{1}{\sqrt{2}}\left( \psi _{1}S(\uparrow )\psi _{2}S(\rightarrow )-\psi _{1}S(\rightarrow )\psi _{2}S(\uparrow )\right) ,
\end{equation}
 where the system's state is supposed to be the difference of permutations of
polarised photon pairs. Since each pair can have one \emph{or} the other orientation-combination
at a time, the summands are logically mutually exclusive. Nevertheless, according
to the `Copenhagen' interpretation, this combination state is considered the
`real' ontological state of the system until measurement `collapses' this wave
function to one or the other `non irreal' summand---as must happen since irreal
states are \emph{never} actually observed in experiments.

Irreality of wave functions is nowadays of relatively low concern; an explanation
of a possible reason for this should emerge below. In part, this is due to the
fact that in many cases the summands are not mutually exclusive and the wave
function exhibits simple, and non problematic, `superposition.'

A natural question here is: just how did this situation arise; what reasoning
lead to accepting such an extraordinary supposition? What problems brought this
reasoning about? To this writer it appears that the answer should be found exactly
there, where the first appearance of wave functions of the form of Eq. (\ref{1})
arose in the literature. Almost certainly, it is in \noun{Heisenberg}'s initial
treatment of the two electron atom, helium.\cite{1} 

His initial efforts to solve this problem were aimed primarily at getting a useful
answer for spectroscopy and only incidentally at developing and promulgating
his preferred paradigm. As a `test bed' for developing the appropriate formalism,
\noun{Heisenberg} chose the problem of coupled harmonic oscillators.\cite{1}
This problem is parallel to the problem of the helium atom in that each electron
is primarily influenced by the nucleus and only secondarily by the other electron,
analogously to oscillators whose behaviour is primarily determined by the `spring
constant(s)' and secondarily by a relatively weak coupling between the oscillators.

\noun{Heisenberg} observed, that it is a characteristic feature of atomic
systems, that the components of which they are comprised, namely electrons,
are identical and subject to identical forces. Therefore, in order to invest
this feature in his `test bed', he assumed the \noun{Hamilton}ian to be of
the form:
\begin{equation}
\label{3}
H=\frac{1}{2m}p^{2}_{1}+\frac{m}{2}\omega ^{2}q_{1}^{2}+\frac{1}{2m}p_{2}^{2}+\frac{m}{2}\omega ^{2}q_{2}^{2}+m\kappa q_{1}q_{2};
\end{equation}
 i.e., the frequencies and masses of the coupled oscillators are taken to be
identical. In Eq. (\ref{3}), \( q_{1},\, q_{2} \) denote the coordinates,
\( p_{1},\, p_{2} \) the momenta, \( m \) and \( \omega  \) the mass and
frequency respectively, and \( \kappa  \) the interaction constant. With help
of the well known transformations:
\begin{equation}
\label{4}
q'_{1}=\frac{1}{\sqrt{2}}(q_{1}+q_{2}),\; \; \; q'_{2}=\frac{1}{\sqrt{2}}(q_{1}-q_{2}),
\end{equation}
 Eq. (\ref{3}) is transformed into the separated form:
\begin{equation}
\label{5}
H=\frac{1}{2m}p_{1}'^{2}+\frac{m}{2}\omega _{1}'^{2}+\frac{1}{2m}p_{2}'^{2}+\frac{m}{2}\omega _{2}'^{2}q_{2}'^{2},
\end{equation}
 where 
\begin{equation}
\label{6}
\omega _{1}'^{2}=\omega ^{2}+\kappa ,\; \; \; \omega _{2}'^{2}=\omega ^{2}-\kappa .
\end{equation}

In other words, \( H \) separates into the sum of two oscillators, such that
each corresponds to a ``normal mode'', in the technique long before developed
by \noun{Daniel Bernoulli}. When only the first mode, \( q'_{1} \), is excited,
then both masses oscillate in phase, and when only \( q'_{2} \) is excited,
out of phase. 
\vspace{0.3cm}

The energies according to QM for the combined system are then give by the equation:
\begin{equation}
\label{7}
H_{n'_{1},n'_{2}}=\frac{\omega '_{1}h}{2\pi }\left( n'_{1}+\frac{1}{2}\right) +\frac{w'_{2}h}{2\pi }\left( n'_{2}+\frac{1}{2}\right) ,
\end{equation}
 where \( n'_{1} \) and \( n'_{2} \) are integers. 

In his scheme, the solutions that \noun{Heisenberg} obtained are matrix elements
found using his version of QM. The solutions from Eq. (\ref{7}) are, as is
usually the case for normal coordinates, not physically observable, but particular
solutions of an abstract combined system. The observables are the inverses of
Eqs. (\ref{4}). At this stage the solutions do not yet suffer irreality; indeed,
the classical mechanical solutions present no philosophical problems. If the
initial conditions are appropriate, the system executes motion described by
one of the normal modes, otherwise, the solution is a secular oscillation of
the total system energy between the two oscillators. 

Observing that, at the atomic scale it is not possible to determine the exact
details of light absorption and emission, \noun{Heisenberg} asserted, not
altogether cogently, that he considered discontinuities more faithful to reality
than \noun{Schr\"odinger}'s \emph{continuous} waves.\cite{2} It is reasonably
arguable, however, that actually he succumbed to sociological pressure, as portrayed
by \noun{Forman}, namely to conform to the pervasive antideterministic philosophical
proclivities prevailing in German academia following World War I.\cite{3} Thus,
with scant underpinning, seemingly in order to accommodate the \emph{Zeitgeist},
he simply \emph{chose} a paradigm involving intrinsic randomness. This, \noun{Heisenberg}
realized by supposing that the solutions, in place of secular oscillation, exhibit
random, spontaneous, secular-like jumping back and forth. 

Instantaneous jumping by itself, is not necessarily
irreal; implicitly there can be a hidden variable that specifies as a
function of time just which electron is excited in the series of jumps back
and forth, that perhaps an extention of QM could predict. However, admitting
this possibility would undermine the sociological goal of discrediting determinism;
and so, for whatever reason, this possibility was rejected out of hand.

The explicit insinuation of the `completion' assumption into the paradigm, or
the `Copenhagen' interpretation, was a complicated and turbid development, the
history of which has been analysed extensively by \noun{Beller}.\cite{4}
Notably, \noun{von Neumann} took up the question of completion at the latest by
1932. He proffered a demonstration to the effect that presuming the existence
of hidden variables completing QM implied that some existing quantum structure is objectively
false.\cite{6} Although this seemed to settle the question, it was quickly seen
(but not widely heeded) that his argument contained irrelevant hypothetical
inputs.\cite{7} In general, both those supporting \noun{Heisenberg}'s
discontinuous and \noun{Schr\"odinger}'s continuous paradigms seemed more eager
than not to assume that quantum theory \emph{is} complete. Presumably, this
happened, to some degree uncritically, as it satisfied the ambition of the
participants to be creating a deep and fundamental new theory; and moreover, it
did not clash with the prevailing cultural bias.

The strictly logical consequence, the implicit paradox, of this assumption, however,
was not assiduously analysed until later after the renowned paper by \noun{Einstein,
Podolsky} and \noun{Rosen} (EPR).\cite{5} It was only with the controversy
evoked by their arguments that the consequences of `completeness', became a
generally acknowledged issue. For example,
\noun{Schr\"odinger} reacted immediately with analysis of the then current
understanding of the meaning of QM in which he introduced the term ``entanglement''
for that form of correlation attributed to irreal wave functions.\cite{9} In
his paper on this matter there are no new quantum techniques introduced, just
new terminology to facilitate deliberate analysis of the then just implicit
connotations for the terms used discussing interpretations. This work, being overtly
critical, was no doubt a contribution to the duel with \noun{Heisenberg} on
the relative merits of discontinuous (matrix) versus continuous wave paradigms.
In it \noun{Schr\"odinger} embellished EPR's illustrative gedanken experiment
to the now renowned and absurdly irreal live-dead ``cat paradox''.

The crucial point here is, if QM is complete, then there can be no hidden variables
to specify which excited state among the constituents at any moment is ontologically
valid, thereby giving their sum this role. That is, then all components, even
mutually exclusive options, are to be extant simultaneously, even when not verifiable
by observation. In short, if QM \emph{is} complete, then there must be \emph{irreal}
states!

\section{Nonlocality}

Nonlocality was introduced as the cure for irreality. The fact that observations
\emph{never} (could!) reveal states that are comprised of irreal sums of mutually
exclusive options, implies, it was hypothesised, that measurement itself somehow
``collapses'' the ontological wave function to the `post-measurement wave function',
that is, just one of the options comprising the irreal, `pre-measurement' wave
function.\footnote{%
Although many early papers read as if their authors imagined wave `collapse'
upon measurement, \noun{von Neumann} appears to have been the first to have
made the matter explicit.\cite{6} The issue is confounded by the fact, that
measurement is to collapse simple superpositions also.
} Insofar as measurement of one of a correlated pair instantaneously collapses
a wave function for the other, regardless of its separation, the process insinuates
`nonlocality'.

\noun{John Bell} in the 1950's\noun{,} having rediscovered \noun{von Neumann}'s
misstep and with inspiration from \noun{Bohm}ian mechanics, took up the issue
with the goal of bringing it to an experimental nexus.\cite{8} He did this
with analysis subsequently, and strictly incorrectly, labelled a ``theorem'',
to the effect that locality demands that a certain statistic (Eq. (\ref{t})
below) be less than \( |2| \). Experiments show, however, that it can reach
\( |2\sqrt{2}| \); and, nowadays the difference is taken to characterise ``stronger
than classical'' correlations which have become denoted ``entanglements''. In
recent times this matter has taken on, so to speak, a life of its own, that
is, irreality has slid into oblivion and usually not discussed as the \emph{raison
d'\^etre} for nonlocality

In any case, \noun{Bell}'s final conclusion was that, because QM is ineluctably
nonlocal, any insinuation of hidden variables to `complete' it, cannot lead
to a deeper formulation that is `local' and `real'. In turn, however, \noun{Bell}'s
argument too has come under criticism, starting with \noun{Edwin Jaynes,\cite{10},}
who parsed \noun{Bell}'s encoding of locality and found that it overlooked
structure requiring \noun{Bayes'} formula for \emph{conditional} probabilities.
This writer has taken up this line and extended it by working out explicit consequences
of \noun{Jaynes}' point for the experiments thought to verify \noun{Bell}'s
analysis.\cite{11,13} That is, classical, local, realist models for all the
generic forms of EPR-type experiments have been developed which lead to calculations,
based essentially on \noun{Malus'} Law, utterly devoid of irreality and nonlocality,
yielding curves precisely mimicking data taken in EPR experiments, \(
|2\sqrt{2}| \)
and all. Since \noun{Bell}'s theorem states in effect that such models do
not exist, exhibiting them shows that \noun{Bell}'s theorem is wrong or misunderstood.

\noun{Jaynes}' essential point is that whereas \noun{Bell} wrote the joint
probability \( P(a,b) \) for a coincidence event in an EPR experiment, where
the measurement settings are \( a \) and \( b \), and \( \lambda  \) represents
the imputed ``hidden variables'', in the form:
\begin{equation}
\label{8}
P(a,b)=\int d\lambda \, \rho (\lambda )P(a,\lambda )P(b,\lambda ),
\end{equation}
 \noun{}he should have written:
\begin{equation}
\label{9}
p(a,b)=\int d\lambda \, \rho (\lambda )P(a|b,\lambda )P(b|\lambda ),
\end{equation}
 where the latter form employs what is known as \noun{Bayes}' formula or simply
as the definition of \emph{conditional} (versus absolute) \emph{}probability\emph{.}\cite{15}
Eq. (\ref{9}), it can be easily verified, does not admit deriving any form
of the renowned ``\noun{Bell} inequalities''. What this means is that \noun{Bell}
misencoded `locality' as `statistical independence', so that the observed violation
of such inequalities in experiments cannot be interpreted to mean that nonlocal
interaction or nonlocal correlation is in evidence. Rather, only, that the inequalities
pertain when there is no correlations of any type, nonlocal or otherwise, contrary
to \noun{EPR'}s, and subsequently to \noun{Bell}'s, hypothesis and the explicit
design of experiments involving \emph{correlated} pairs of inputs. 

Since this matter has been explicated in detail elsewhere, \cite{11,13,14},
here only one variant of several counter arguments shall be featured. Its key
idea is that if the physical meaning of the terms in \noun{Bell}'s extraction
of his inequalities are carefully interpreted, it is seen that certain of them
must be zero, thereby leading to a form of these inequalities for which there
is no significance with regard to his sought after conclusion. This counter
argument, which is independent of \noun{Jaynes'} criticism, but based on the
same structure, proceeds as follows:

First, recall a mathematical technicality concerning the product of two Dirac
delta functions, which is essential for what follows. It is that the integral
of the product of two delta functions for which the arguments are different,
equals zero; i.e.: 
\begin{equation}
\label{aa}
\int dx\, f(x)\delta (x-l)\delta (x-m)=0,
\end{equation}
 whenever \( l\neq m \).

The derivation of a Bell Inequality starts from \textsc{Bell}'s fundamental
assertion:
\begin{equation}
\label{p}
P(a,\, b)=\int d\lambda \, \rho (\lambda )A(a,\, \lambda )B(b,\, \lambda ),
\end{equation}
 where, per \emph{explicit assumption:} \( A \) is not a function of \( b \);
nor \( B \) of \( a \); and each represents the appearance of a photoelectron
in its wing of an EPR experiment, and \( a \) and \( b \) are the corresponding
polariser filter settings.\(^2\)
This is motivated on the grounds that a measurement at station \( A \), if
it respects `locality', \emph{so argues} \textsc{Bell}, cannot depend instantaneously
on remote conditions, such as the settings of the other polariser. In addition,
each, by definition, satisfies
\begin{equation}
\label{pp}
|A|\leq 1,\; \; |B|\leq 1,
\end{equation}
 which in this case effectively restricts the analysis to the case of just one
photoelectron per time window per detector. Eq. (\ref{p}) encodes the condition,
that when the hidden variables are averaged out, the usual results from QM are
to be recovered.

The \( \lambda  \) above in \textsc{Bell}'s analysis stands for a hypothetical
set of ``hidden variables'', which, if they exist, should render QM deterministic.
This set may include many different types of variables, such as discrete, continuous,
tensor or whatever.

\begin{widetext}

Extraction of inequalities proceeds by considering differences of two such correlations
where \( (a,\, b) \), \emph{i.e.}, the polariser axis of measuring stations,
left and right, differ:
\begin{equation}
\label{q}
\begin{array}{l}
P(a,\, b)-P(a,\, b')=\int d\lambda \, \rho (\lambda )[A(a,\, \lambda )B(b,\, \lambda )-A(a,\, \lambda )B(b',\, \lambda )],
\end{array}
\end{equation}
 to which zero in the form:
\begin{equation}
\label{r}
\begin{array}{l}
A(a,\, \lambda )B(b,\, \lambda )A(a',\, \lambda )B(b',\, \lambda )-A(a,\, \lambda )B(b',\, \lambda )A(a',\, \lambda )B(b,\, \lambda )=0,
\end{array}
\end{equation}
 is added to get: 
\begin{equation}
\label{s}
%\begin{array}{l}
P(a,\, b)-P(a,\, b')=\int d\lambda \, \rho (\lambda )A(a,\, \lambda )B(b,\,
\lambda )[1\pm A(a',\, \lambda )B(b',\, \lambda )]
%\\ \phantom{P(a,bc)-xxxxPa,cf)}
-\int d\lambda \, \rho (\lambda )A(a,\, \lambda )B(b',\,
\lambda )[1\pm A(a',\, \lambda )B(b,\, \lambda )],
%\end{array}
\end{equation}
 which, in turn, upon taking absolute values and in view of Eqs. (\ref{pp}),
\textsc{Bell} wrote as:
\begin{equation}
\label{ss}
\begin{array}{l}
|P(a,\, b)-P(a,\, b')|\leq \int d\lambda \, \rho (\lambda )[1\pm A(a',\, \lambda )B(b',\, \lambda )]+\int d\lambda \, \rho (\lambda )[1\pm A(a',\, \lambda )B(b,\, \lambda )].
\end{array}
\end{equation}
 Then, using Eq. (\ref{p}), and the normalisation condition \( \int d\lambda \, \rho (\lambda )=1, \)
he got, for example:
\begin{equation}
\label{t}
|P(a,\, b)-P(a,\, b')|+|P(a',\, b')+P(a',\, b)|\leq 2,
\end{equation}
 a `Bell inequality'.
 
\end{widetext} 

Now, however, if the \( \lambda  \) are a complete set
\setcounter{footnote}{1}\protect{\footnote{%
\noun{Bell}'s notation, e.g., \( P(a,\lambda ) \), makes no distinction between
variables, here \( a \), and conditioning parameters, here \( \lambda  \),
customarily separated by a vertical bar rather than a comma. This oversight
is the source of much confusion, and possibly even the subliminal cause of his
`error'. In this paper, Bell's notation is retained whenever referring directly
to his formulas.}}\(^,\)\setcounter{footnote}{2}\footnote{%
\noun{Bell} used a single symbol: \( \lambda  \), to denote what could be
a complicated set of variables of possibly different types even. Thus, a ``particular
values for \( \lambda  \)'' means that each entity in the whole set must have
a value.
}, thereby rendering everything deterministic so that all probabilities as functions
of \( \lambda  \) become Dirac or Kronecker delta distributions, then the \( A \)'s
and \( B \)'s in Eq. (\ref{s}) are pair-wise; that is to say as individual
events comprising the generation at the source of one pair, are non zero for
distinct values of \( \lambda  \), which, by virtue of completeness, do not
coincide for distinct events, \emph{i.e.}, for different pairs. That is, for
each pair of settings \( (a=r,\: b=s) \) and iteration of the experiment, \( n \),
there exists a unique set of values, \( \lambda _{a=r,\, b=s}(n=\textrm{integer}) \),
or in more compact notation, \( \lambda (n) \), say, for which \( A(a|\lambda (n))B(b|\lambda (n)) \)
is non-zero (\( 1 \) in the discrete case, \( \infty  \) in the continuous
case). In other words, each product \( A(a|\lambda (n))B(b|\lambda (n)) \)
can be written in the form \( f(x)\delta (x-\lambda (n)) \), so that all quadruple
products, e.g.,
\begin{equation}
\label{u}
A(a|\lambda (n))B(b|\lambda (n))A(a|\lambda (m))B(b|\lambda (m)),
\end{equation}
 are equivalent to the form: 
\begin{equation}
\label{v}
f(x)\delta (x-\lambda (n))g(x)\delta (x-\lambda (m)),
\end{equation}
 where \( x \) is a dummy variable of integration to run over all admissible
values of \( \lambda  \). Therefore, such terms with pair-wise different values
of \( \lambda (n) \) in Eq. (\ref{s}), \emph{i.e.}, whenever \( n\neq m \)
, are, in accord with Eq. (\ref{aa}), identically zero under integration over
\( \lambda  \). This annihilates two terms on the right side of Eq. (\ref{s}),
so that the final form of this Bell Inequality is then actually the trivial
identity: 
\begin{equation}
\label{gg}
|P(a,\, b)|+|P(a',\, b')|\leq2 .
\end{equation}

Thus, our final conclusion is, that the proof of the ineluctability of the presence
of nonlocality in QM, is invalid. In the context of what is actually a further
point regarding the admissibility of additional, ostensibly still `hidden' variable
completions of QM, this conclusion undermines the popular impression that such
a `completion' necessarily cannot reinstate `reality' and `locality.'\setcounter{footnote}{3}\footnote{%
\noun{Kochen-Specker} type `no-go' theorems without probabilities also can
be shown to be defective as all their hypothetical inputs cannot be true simultaneously
on physical grounds.\cite{15a}
}

\section{Bohm's version of EPR experiments}

EPR proposed a gedanken experiment involving the disintegration of a mother
particle into two daughters moving off in opposite directions.\cite{5} They
observed that quantum principles state that \noun{Hamilton}ian conjugate variables
suffer \noun{Heisenberg} uncertainty, and therefore, cannot be simultaneously
determined exactly. But, in the situation envisioned by EPR, the experimenter
could observe one daughter's position to arbitrarily high precision, while observing
to arbitrarily high precision the other daughter's momentum; and, then, calling
on symmetry, one could specify to arbitrary high precision all four quantities---in
conflict with \noun{Heisenberg}'s Principle. Hence, the paradox.

Complicating matters, actually doing the EPR experiment as proposed, is impractical.
So, as is well known, \noun{Bohm} proposed another venue, now called ``qubit''
space.\cite{16} Originally he considered using particles with spin; but, this
venue too was impractical, so experimenters chose spin's homeomorphic partner:
light polarisation space. This change of venue, however, introduces two serious
defects. One, the two states of polarisation, in spite of not commuting (for
purely geometric reasons it turns out), are not \noun{Hamilton}ian conjugate
pairs; the structure involved in not intrinsic to QM, but just to classical
electrodynamics. Two, the two polarisation states are non interacting; the structure
leading to `irreal' states as the sum of two mutually contradictory summands,
is not relevant. Appropriate states for emissions of polarisation correlated
daughter signals can be expressed without use of the `irreal' format. In conclusion,
even disregarding \noun{Jaynes'} challenge to their general validity, various
hypothetical inputs into the logic of the derivation of \noun{Bell} inequalities
have not been met by optical experiments widely credited with ``proving \noun{Bell}'s
theorem''.\setcounter{footnote}{4}\footnote{%
EPR experiments on particles have given largely ambiguous results for additional,
independent reasons. See: \cite{17} for a current review of EPR experiments,
albeit without acknowledging \noun{Jaynes'} arguments.
}

\section{Summary and forecast}

Herein, history has been, so to say, ``rewritten'' so as to parse better the
logical interrelationships among the features constraining development of an
interpretation for QM. The concluding point here is that the assumption that
quantum theory is ``complete'' compelled the introduction of the notions of
`irreality' and `nonlocality'. That the assumption of `completeness' is the
source of conflicted logic is by far not here unique; \noun{Post} has argued
convincingly already for decades, on the basis of detailed analysis of numerous
specific ``quantum'' phenomena, that this assumption introduces serious lacuna.\cite{18} 

Unfortunately, simply rejecting the completeness assumption alone does not render
the matter clear. QM captures an undulatory feature even deeper than the structure
involving the statistics of ensembles of atomic scale entities; this requires
a physical nature for `wave functions'; despite \noun{Schr\"odinger}'s failure
to find it, they cannot simply be just progenitors of statistical densities.
A full remedy must be found elsewhere.\setcounter{footnote}{5}\footnote{%
Not surprisingly, this writer prefers his own proposal for a resolution of the
enigma of the origin of wavelike probabilistic features in QM.\cite{19}
} 

Independently and additionally, I argue in support of \noun{Jaynes,} that
\noun{Bell}'s analysis stemming from the EPR arguments in favour of the incompleteness
of QM, contains an error. 

Beyond the philosophical (or accompanying mathematical) implications of revamping
the nowadays customary understanding of this issue in terms of ``entanglement'',
there is a possible solidly practical consequence. It follows from the fact
that the physical manifestations of entanglement targeted for exploitation,
all depend in the end on the consequences of superposition resulting from the
linearity of the hyperbolic differential equation underpinning QM, i.e., \noun{Schr\"odinger}'s
equation. Since this hyperbolic structure is not an exclusive consequence of
any quantum feature, certain applications now thought to require atomic scale
realization, `quantum logic gates' for example, may in fact not need micro devices
to run what are thought to be `quantum algorithms'. As manufacturing macroscopic
devices is much easier, a computer realizing the parallelism of what turns out
to be implicit in all \noun{Fourier} analysis, not just from `quantum entanglement',
might be eminently obtainable.\setcounter{footnote}{6}\footnote{%
This speculation assumes, of course, the physical cogency of `quantum algorithms',
a matter on which this writer holds no professional opinion as of yet. But there
is room for optimism if these algorithms require only superposition and not
irreality.
}

\end{document}